\documentclass[pdflatex,sn-mathphys-num]{sn-jnl}

\usepackage{graphicx}%
\usepackage{multirow}%
\usepackage{amsmath,amssymb,amsfonts}%
\usepackage{amsthm}%
\usepackage{mathrsfs}%
\usepackage[title]{appendix}%
\usepackage{xcolor}%
\usepackage{textcomp}%
\usepackage{manyfoot}%
\usepackage{booktabs}%
\usepackage{algorithm}%
\usepackage{algorithmicx}%
\usepackage{algpseudocode}%
\usepackage{listings}%
\usepackage{indentfirst}
\theoremstyle{thmstyleone}%
\theoremstyle{thmstyletwo}%

\theoremstyle{thmstylethree}%
\newcommand{\CTS}{Co$_{1/3}$TaS$_2$} 

\begin{document}

\title[Article Title]{Tunable decoupling of coexisting magnetic orders in \CTS}


\author*[1]{\fnm{Yining} \sur{Hu}}\email{yining.hu@mpsd.mpg.de}

\author[2,3]{\fnm{Zili} \sur{Feng}}\email{zlfeng@caltech.edu}

\author[2,3]{\fnm{Takashi} \sur{Kurumaji} }\email{kurumaji@caltech.edu}

\author*[2,3]{\fnm{Linda} \sur{Ye}}\email{lindaye@caltech.edu}

\author*[1]{\fnm{Chunyu Mark} \sur{Guo}}\email{chunyu.guo@mpsd.mpg.de}

\author*[1]{\fnm{Philip J. W.} \sur{Moll}}\email{philip.moll@mpsd.mpg.de}

\affil[1]{\orgname{Max Planck Institute for the Structure and Dynamics of Matter}, \orgaddress{\city{Hamburg}, \country{Germany}}}

\affil[2]{\orgdiv{Division of Physics, Mathematics and Astronomy}, \orgname{California Institute of Technology}, \orgaddress{\city{Pasadena}, \state{CA} \postcode{91125},  \country{USA}}}

\affil[3]{\orgdiv{
Institute of Quantum Information and Matter}, \orgname{ California Institute of Technology}, \orgaddress{\city{Pasadena}, \state{CA} \postcode{91125},  \country{USA}}}

\abstract{In multiferroics, new physical responses and functionalities emerge when symmetry-distinct order parameters couple. This conventionally occurs when lattice and magnetic degrees of freedom order independently in a material. Here, we report an all-magnetic analogue of multiferroic behavior in the antiferromagnet \CTS, where topological scalar spin chirality and nematicity coexist on the same spin lattice. While the chiral spin texture generates an anomalous Hall effect (AHE), the nematic order breaks threefold rotational symmetry and dominates longitudinal transport. Crucially, in zero field these symmetry-distinct orders merely coexist yet magnetic fields induce strong coupling between them, thus realizing a new type of multiferroic bebhavior via tuning of the coupling itself instead of direct manipulation of secondary orders. In sub-domain sized devices with achiral geometry, we demonstrate that nonreciprocal transport serves as a symmetry-based probe of the global spin chirality, co-aligned with the strong topological AHE of the system. In \CTS~the topological Hall state inherits a large resistance anomaly via chiral-nematic coupling, thus our results showcase how hybrid magnetic orders can achieve advanced functionalities by merging symmetry-forbidden material responses.
}



\maketitle 

\section*{Main}

\begin{figure}[H]
    \centering
    	\includegraphics[width = 0.98\linewidth]{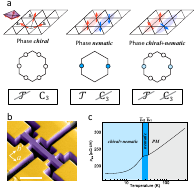}
\caption{\textbf{a} The real-space and reciprocal-space sketch of chiral phse, nematic phase and their combination. The volume spanned by Si, Sj and Sk is propotional to the chirality. The time-reversal symmetry and the C$_{3z}$ rotational symmetry of the three phases are indicated. \textbf{b} SEM image of the device for transport measurement the $b'$ axis denotes the crystallographic $b$ axis rotated by $30^\circ$ clockwise in device plane. The current is applied along $a$ direction. Scale bar 10 $\mu$m. \textbf{c} Temperature dependence of longitudinal resistivity.}
    \label{fig1}
\end{figure}

When magnetic materials order below a transition temperature, macroscopic numbers of microscopic degrees of freedom condense into a collective state described by an order parameter with only a few components. This reduction both poses the conceptual challenge of identifying and classifying often complex magnetic orders and creates opportunities for applications, as the discrete states of an order parameter can be used to store or process information. 

The crux of exploiting ordered states is always a balance between mutually incompatible requirements. On the one hand, the order should be robust against external perturbations to ensure stability and low noise; on the other hand, it must remain susceptible to controlled manipulation by weak stimuli, such as low-power electrical currents. Identifying physical mechanisms that satisfy both requirements remains a central challenge. One successful paradigm is provided by multiferroics \cite{SpaldinFiebig2005,SpaldinPhysToday2010}, in which multiple coupled order parameters coexist, allowing one form of order to be manipulated indirectly by stimulating another. Representative examples span the type-I multiferroics such as BiFeO$_3$ \cite{WangBiFeO32003} where electric fields control magnetic domains via ferroelectric switching \cite{HeronNature2014,Zhao2006_BiFeO3_AFMdomains}, and type-II multiferroics such as TbMnO$_3$ \cite{PhysRevLett.98.147204} and MnWO$_4$ \cite{PhysRevB.81.054430} where electric fields select the handness of spiral magnetic order. In such systems, symmetry-allowed coupling terms between distinct orders are set by the crystal structure. 

Here we report an all-magnetic analogue in the antiferromagnet \CTS~\cite{ParkinFriend1980MagProps,ParkinFriend1980Transport,ParkinMarsegliaBrown1983JPhysC,ParkinFriend1980PhysicaBC}. At low temperatures, the Co moments participate in two distinct and symmetry-incompatible antiferromagnetic (AFM) orders \cite{Park2023}, associated with the breaking of time-reversal and threefold rotational symmetries, respectively (Fig. 1a). Each order produces a qualitatively different signature in electronic transport. One corresponds to a chiral, topological AFM state with a pronounced anomalous Hall response, which arises from a finite scalar spin chirality \cite{takagi2023,Park2023} set by the volume spanned by a non-coplanar spin texture (Fig. 1a). Meanwhile, the other is a nematic AFM state that strongly affects the longitudinal resistance \cite{feng2025,kirstein2025}. These orders coexist at low temperatures yielding a both chiral and nematic state, yet their incompatible symmetries prevent a direct coupling between them. We demonstrate, however, that out-of-plane magnetic fields induce and control their coupling, driving a transition between decoupled coexisting towards strongly intertwined orders by the application of a magnetic field.

It presents two main advances in our understanding and control of complex magnets. First, it demonstrates how inducing symmetry-forbidden coupling between orders allows to dress and undress magnetic states endowing them with new functionalities and properties in a controlled way. For example, orders may be selectively coupled for efficient write operations, and then decoupled to protect the information from perturbations. Here specifically the topological chiral order, which by itself has little influence on the resistance, stabilizes the coexisting nematic state and endows its repolarization with a substantial change in resistance. As a result, the sign of the chiral order can be read out through a simple high-resistance/low-resistance measurement, while removing the magnetic field decouples the states again.

Second, the chiral magnetic texture gives rise to a pronounced second-harmonic signal in longitudinal transport currents that is directly proportional to the spin chirality, a phenomenon known as electromagnetic chiral anisotropy (eMChA) \cite{PhysRevB.99.245153, PhysRevLett.87.236602, PhysRevLett.95.256601, PhysRevLett.122.057206, Yokouchi2017ElectricalMagnetochiral, Pop2014ElectricalMagnetochiral, Guo2022SwitchableChiral}. This unambiguously confirms the connection between the anomalous Hall effect and the scalar spin chirality, as has been theoretically predicted \cite{NagaosaAHE2010}. Together, our results demonstrate a new control paradigm for complex magnets, in which external fields tune not only order parameters but their mutual coupling, enabling switchable topological–nematic states with directly accessible transport functionality.

\CTS~is a Co-intercalated transition-metal dichalcogenide (TMD) in which magnetic $\mathrm{Co^{2+}}$ ions ($S~\mathrm{\backsim1.35}$ from Curie-Weiss result \cite{Park2023}) form triangular layers in the van der Waals gaps, and the remarkably rich sequence of ordered states in this metallic frustrated AFM has been studied intensely \cite{ParkinFriend1980MagProps,ParkinFriend1980Transport,ParkinFriend1980PhysicaBC,ParkinMarsegliaBrown1983JPhysC}
. More recently, \CTS~attracted broad interest for its large anomalous Hall response below the transition temperature $(T_{N2}\approx 26.5~\mathrm{K} )$ despite its negligible net magnetization ($10^{-3}\mu_B/Co$), placing it among the key experimental platforms for topological anomalous Hall physics in non-collinear AFM \cite{Park2023,takagi2023}. While its complex magnetic structure remains under debate \cite{Park2022,Park2023,takagi2023,ParkComposition2024,kirstein2025,feng2025,ParkSpindynamic2025}, the central microscopic ingredient generally invoked in this low-temperature regime is a non-coplanar spin texture (often discussed as a multi-$Q$ state) that generates an emergent real-space Berry curvature (Fig.1). Time-reversal symmetry is broken by a finite scalar spin chirality on spin plaquettes ($\chi=\mathbf S_i\cdot(\mathbf S_j\times \mathbf S_k)$). Akin to a $\mathbb{Z}_2$ Ising magnet, the effective gauge field of the spin chirality can point in two directions, up or down. The threefold rotational symmetry $C_3$  in the pure chiral AFM phase (white region in Fig. 2a), however, is preserved as evidenced by the absence of linear dichroism \cite{kirstein2025} and in-plane resistivity anisotropy \cite{feng2025}.

In addition to the chiral AFM low-temperature phase, experiments establish a second ordered state emerging below $T_{\mathrm{N1}} \approx 38~\mathrm{K}$ that is distinct in both symmetry and transport response. This phase is a collinear, single-$Q$ antiferromagnetic stripe order that breaks the underlying threefold rotational symmetry and is therefore commonly discussed as a nematic antiferromagnetic order. For brevity, we refer to this single-$Q$ order as nematic despite it breaking translational symmetry. The rotational symmetry breaking is directly observed via linear dichroism \cite{kirstein2025} and a strong in-plane resistivity anisotropy \cite{feng2025}, while the absence of an anomalous Hall effect and circular dichroism establishes that this state preserves time-reversal symmetry. The nematic and chiral phases thus belong to symmetry-incompatible channels.

These recent studies of dichroism further substantiate the absence of coupling between the two orders \cite{kirstein2025}. Circular and linear dichroism simultaneously image the rich domain patterns of both orders, yet without domain correlation confirming their independent coexistence in zero field. To access these orders and their coupling in transport experiments, the material should be probed at a single domain length scale. Hence micron-sized bars aligned along the crystallographic $a$-direction were fabricated by focused ion beam machining (Fig. 1b) with device dimensions comparable to the typical size of chiral AFM domains and substantially smaller than that of nematic domains in bulk crystals \cite{kirstein2025}. The microstructures further support large current densities facilitating non-linear measurements compared to macroscopic bulk crystals.

Both the temperature- and field-dependent resistivity measured in all five microstructures reproduces the key features reported previously in bulk samples  \cite{Park2022,feng2025} with no substantive microstructure-induced differences. The consistent behaviour across devices indicates good reproducibility (see supplement Fig. S1-S3). The temperature dependence of the longitudinal resistivity $\rho_{aa}(T)$ shows a sharp cusp at $T_{N1}$, signaling the nematic transition, whereas the onset of the chiral phase below $T_{N2}$ produces only a broad rollover, demonstrating that longitudinal transport is dominated by nematic order.

\begin{figure}[H]
    \centering
    	\includegraphics[width = 0.98\linewidth]{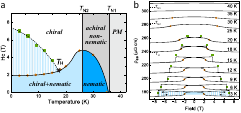}
\caption{\textbf{a} The phase diagram of \CTS. The area with blue stripes is the new phase defined by our experiment, which is pure chiral phase or chiral+nematic phase. \textbf{b} Magnetoresistance at different temperatures. The orange circles denote the metamagnetic transition for nematicity, and the green squares denote the chirality reversal transitions.}
    \label{fig2}
\end{figure}

Applying an out-of-plane magnetic field reveals a rich phase diagram (Fig. 2). In the paramagnetic (PM) regime, a weak negative magnetoresistance reflects dominant spin-disorder scattering. For the purely nematic phase below $T_{N1}$ above $T_{N2}$, a pronounced inflection point (orange circles in Fig. 2b) appears in the magnetoresistivity, signaling a metamagnetic transition into a field-stabilized state that is both rotationally and time-reversal symmetric as suggested from dichroism. This feature persists smoothly across $T_{N2}$ , indicating that the emergence of chiral order does not alter the field evolution of the nematic state. Just below $T_{N2}$, the high-field phase is a purely chiral state with strong anomalous Hall and circular dichroism signals but no nematicity, while only a very weak hysteretic anomaly is visible in the longitudinal resistance.

Upon further cooling, the coercive field of the chiral state increases and exceeds the metamagnetic transition field at an intermediate temperature $T_H$. Below $T_H$, a wide hysteretic region opens in the phase diagram, accompanied by a qualitative change in the hysteresis loop shape. While upon lowering the field strength the metamagnetic transition occurs unimpeded (Fig. 2b, orange circle), its signature on the upsweep is fully absent. Instead, the system remains in a high resistance state up to the coercitive field (green square), at which a remarkably large resistance drop back into a low resistance state is observed. As the nematic transition yields the strong resistive anomaly and the chirality reversal causes the large coercivity, the fact that both occur simultaneously directly evidences their strong coupling.

\begin{figure}[H]
    \centering
    	\includegraphics[width = 0.98\linewidth]{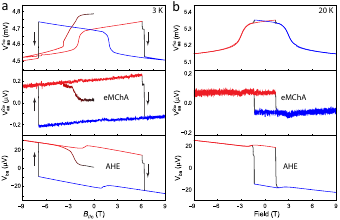}
    \caption{Longitudinal voltage and its second harmonics (eMChA) and the Hall voltage measured at \textbf{a} 3 K and \textbf{b} 20 K respectively. The 20 K eMChA is calibrated by subtracting a constant background induced by heating (see supplement \textbf{Thermal effect on $V_\mathrm{2\omega}$}).}
    \label{fig3}
\end{figure}

While these states appear intertwined in linear transport, they are readily disentangled at the non-linear level (Fig. 3). At 3K, the resistive hysteresis loop is accompanied by a large AHE as reported previously \cite{takagi2023,Park2023}, indicative of the artificial gauge field emerging from the topological spin texture. While commonly associated to a finite scalar spin chirality, the role of finite chirality in low-frequency transport was unclear. In the microbars, large enough current densities can be reached to detect second-harmonic conversion, allowing finite eMChA ($V_{2\omega}$) to be resolved as a first direct transport signature of spin chirality. 

At a given temperature, all three transport signals measured simultaneously undergo an abrupt change at the same magnetic field upon sweeping from zero to high field, indicating a coincident reversal of spin chirality, time reversal, and collapse of nematic order. While the longitudinal resistance reflects the nematic component of the spin texture, the eMChA and the AHE are governed by the breaking of time-reversal symmetry. The transition corresponds to a switch between two noncoplanar spin textures of opposite scalar chirality, related by a time-reversal operation, which inverts the emergent Berry curvature and hence both eMChA and the AHE. Initially, after a zero-field cooldown the system is in a multidomain state, as evidenced by the absence of AHE and eMChA as well as the excess linear resistance. Field polarizing to about 5T induces a mono-domain state, followed by bi-stable hard-magnet behaviour in subsequent field sweeps.

Upon increasing the temperature across $T_\mathrm{H}$ up to 20 K, the magnetoresistance undergoes a pronounced change in shape, whereas both the eMChA and the AHE retain a simple hysteretic field dependence, with only a reduced coercive field due to the elevated temperature. As a consequence, the chirality reversal shifts to fields lower than the metamagnetic transition, rendering the associated nematic transition non-hysteretic. Consistent with this picture, the magnetoresistance exhibits only a weak response to the chirality reversal, with the corresponding jump reduced by an order of magnitude compared to that at 3 K. 

Viewing the magnetic state evolution through these three distinct transport signals allows to disentangle the role of the different orders. Although the eMChA and the AHE follow the same overall field dependence, a clear deviation between the two responses appears at the nematic transition. The AHE senses both the emergent gauge field of the topological order as well as the ferromagnetic net moment, and hence the nematic transition is clearly visible in the data. eMChA, on the contrary, senses purely the chiral part of the magnetic order. This provides direct evidence that the chiral topological order persists unimpeded by and independent of the nematic one.

This counterintuitive robustness of the scalar spin chirality can be motivated by spin-model calculations, with the main physics sketched in Fig. 4a. Out-of-plane magnetic fields tilt the spins towards the field, which well reduces the solid angle spanned by the three spins of a plaquette. However, the same spins contribute to the neighboring plaquette on the Kagome lattice, and this compensation keeps the average scalar spin chirality constant to first order in applied field.

\begin{figure}[H]
    \centering
    	\includegraphics[width = 0.98\linewidth]{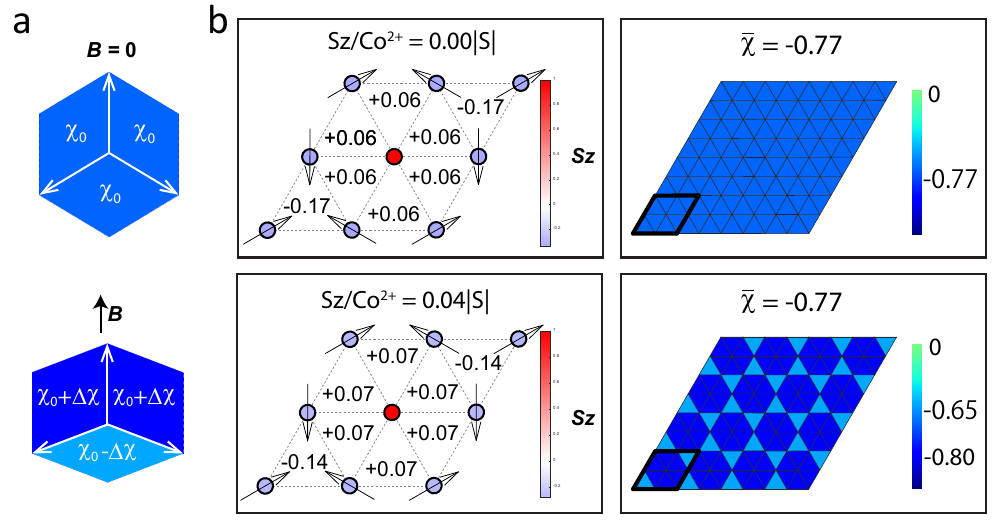}
    \caption{\textbf{a} 2D analogue of the spin texture and the chirality defined by it. \textbf{b} Net magnetization and scalar spin chirality with and without an external magnetic field. In the spin model for net magnetization, the arrows represent the in-plane component of the spins and the color in circles represents the out-of-plane component of the spins. The local out-of-plane magnetization $S^{z}_{\mathrm{local}}$ and the local scalar chirality $\chi_{\mathrm{local}}$ are indicated by the values and colors on each triangular plaquette, respectively. The parameters used in the simulation are $J_1 = 1$, $J_2 = 0.5$, $D = 0.05$, $B = 0.25$ and $K = 0.01$. For simulations with an external field, we choose $H = 0.3$, corresponding to a similar energy scale as in the experiment, estimated using data from Ref.~\cite{Park2023}}
    \label{fig4}
\end{figure}

A more rigorous simulation (Fig. 4b, see supplement) shows that in the absence of an external field, the four spins adopt polar angles of $0^\circ$ or approximately $109^\circ$, forming a configuration close to an ideal equilateral tetrahedron. In this geometry, the net magnetization is nearly compensated, as $\cos(109^\circ)$ is close to $-1/3$, balancing the contribution from the upright spin. Because the angles between any pair of spins are identical, the scalar spin chirality is homogeneous across all triangular plaquettes.

When an external magnetic field is applied along the $z$ direction, the three canted spins tilt upward simultaneously to reduce the Zeeman energy which increases net magnetization, while the spin aligned along $z$ remains essentially unchanged. The scalar spin chirality, however, evolves differently. The triangular plaquettes formed by the three canted spins exhibit a reduced chirality (in amplitude), as the volume spanned by these spins decreases. In contrast, plaquettes formed by two canted spins and one upright spin show an enhanced chirality. As a result, these opposing changes largely compensate, and the average chirality remains nearly unchanged. This contrast naturally explains why the eMChA and the AHE can deviate from each other under applied magnetic fields.

\begin{figure}[H]
    \centering
    	\includegraphics[width = 0.98\linewidth]{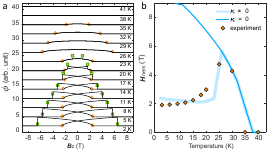}
    \caption{\textbf{a} Calculated nematic order parameter $\phi$ with the field-induced coupling ($\kappa\neq 0$)  to the chiral order parameter $\chi$. The nematic transitions are marked by empty orange circles and the chirality reversal transitions are marked by empty green squares. \textbf{b} Temperature dependence of the nematic transition field $H_\mathrm{nem}$ extracted from experiment (solid circles) and calculations (open circles).}
    \label{fig5}
\end{figure}

\section*{Discussions}

These experiments clearly demonstrate the coexistence and field-induced coupling of two symmetry-inequivalent magnetic order parameters. The physics of such coupled chiral ($\chi$) and nematic ($\phi$) orders can be further elucidated by a phenomenological Landau model (see supplement). The essence is found in the lowest order symmetry-allowed coupling $\propto H_z \chi \phi^2$. Crucially, $\chi$ may enter linearly only in a bilinear form with the out-of-plane magnetic field $H_z$ due to time-reversal symmetry of the free energy. This term, however, allows the nematic order to sense the sign of the chirality given the polarity of an applied field, and is key to reproduce discontinuous switching behavior. In contrast, the rotational symmetry breaking of the nematic order allows only $\phi^2$ as minimal coupling. As a result, the existence or absence of nematicity at a given field depends on the chirality of the magnetic texture. When the chiral order reverses at its coercive field, the sign change of $H\chi$ can abruptly collapse the nematic order. Interestingly, the same coupling coefficient has been found to be crucial to understand the coupling between a rotational-symmetry breaking bond charge order and a time-reversal symmetry breaking loop-current order in the Kagome superconductor CsV$_3$Sb$_5$ \cite{Guo2024CsV3Sb5,PhysRevB.108.125136,Scammell2023ChiralExcitonic,PhysRevB.106.144504,fernandes2025loopcurrentorderkagomelooking,ingham2025vestigialorderexcitonicmother}. This coefficient is uniquely allowed for $C_3$ symmetry breaking planar models and appears to be a universal root of coupled orders in kagome materials. This toy model thus illustrates how a symmetry-allowed Landau coupling enables indirect electrical readout of a topological antiferromagnetic order through its impact on a coexisting nematic phase.

Indeed, this simple Landau model for $\phi$ (Fig.~5a) reproduces the main trends in the magnetoresistance strikingly well (Fig.~2b). The field suppressing nematicity $(\phi(H_{\mathrm{nem}})=0)$ initially increases upon cooling, but is subsequently reduced once the chiral order develops. This reduction indicates that a finite spin-chiral order parameter $\chi$ suppresses $\phi$ (and thus lowers $H_{\mathrm{nem}}$) without eliminating the nematic order.

These observations establish an interesting paradigm in the manipulation of magnetic orders. Instead of manipulating directly or indirectly through rigidly coupled systems as in multiferroics, symmetry-breaking stimuli can instead control the coupling between them. It would be interesting to spatially image this coupling. For instance, after zero-field cooling, the two orders are expected to form independent domain patterns, whereas repeated field-training cycles would induce a spatial correlation between their domain distributions. Further studies on the combined role of electric and magnetic field in the coupling are required, especially in light of recent reversible switching of the scalar spin chirality in \CTS~nanoflakes~\cite{xiong2025allelectricalselfswitchingvander}. 

In practical memory, it is desirable to combine two functionalities: a state that is easily tunable (for writing) and a state that is robust against perturbations (for retention). These requirements correspond to opposing energy-landscape criteria: fast switching favours a low barrier between states, whereas long-term stability requires a high barrier. Here, symmetry-incompatible orders connected by a tunable coupling provide an elegant solution. 

\backmatter

\addcontentsline{toc}{section}{References}
\bibliography{sn-bibliography.bib}

\section*{Methods}

\textbf{Crystal synthesis}\\
Single crystals of \CTS~were grown using a two-step procedure detailed in Ref. \cite{feng2025}. Polycrystalline \CTS~was first synthesized via solid-state reaction from stoichiometric mixtures of high-purity Co (99.99\%, Strem Chemicals), Ta (99.98\%, Strem Chemicals), and S (99.9995\%, Alfa Aesar) powders. The mixed powders were placed in an alumina crucible, sealed in an evacuated quartz ampoule, heated slowly to $900^\circ \mathrm{C}$, and held at that temperature for 24 hours.
Single crystals were subsequently obtained through chemical vapor transport using I$_2$ as the transport agent. The source and sink temperatures were maintained at $950^\circ \mathrm{C}$ and $850^\circ \mathrm{C}$, respectively. After approximately one week, plate-like crystals with lateral dimensions up to 2 mm were harvested. Phase purity was confirmed by powder X-ray diffraction.\\

\textbf{Device fabrication}\\
\CTS~microstructures were prepared using a Thermo Fisher Helios Ga FIB. Lamella for the five devices studied were extracted from the bulk crystal and transferred onto an araldite AB glue droplet on a prepatterned sapphire chip. To form good electrical contact between the lamella and the electrodes on the sapphire chip, the lamellae together with the sapphire chip were treated with radio-frequency (RF) Ar plasma etching for removing the native surface oxide, followed by high-power Au sputtering in the same chamber. After Au deposition, the lamella were patterned into the desired device geometries. \\

\textbf{Electrical measurements}\\
Electrical resistance was measured in a DynaCool PPMS using a multichannel SynkTek lock-in amplifier, with magnetic fields up to 9 T. For all devices, an AC excitation current from $400 ~\mathrm{\mu A}$ to $1470 ~\mathrm{\mu A}$ was applied by Keithley 6221 current source. Phase lock loop (PLL) is used to lock the demodulator to the excitation reference, so that it can track the frequency/phase precisely. The resistivity is read out by the first-order harmonic channel and the eMChA is read out by the second-order harmonic channel. Across the full device set, clear MR and eMChA were consistently observed, indicating that the FIB-fabricated microstructures retain their intrinsic transport properties. The FIB-induced amorphous surface layer, typically around 20 nm thick, therefore represents a negligible fraction of the device cross-section (except for device S4) and does not affect the reported measurements.\\

\section*{Data availability}
The data supporting the findings of this study will be deposited on Zenodo.\\

\section*{Code availability}
The code supporting the findings of this study will be deposited on Zenodo.\\

\section*{Acknowledgements}
C. M. G. acknowledges financial support by the European Research Council (ERC) under grant Free-Kagome (Grant Agreement No. 101164280). The material synthesis efforts carried out at Caltech are supported by Gordon and Betty Moore Foundation through Moore Materials Synthesis Fellowship to L.Y. (GBMF12765).  Z.F. acknowledges support from Quantum Information and Matter (IQIM) Postdoctoral Fellowship at Caltech. We thank K. Wang for discussions related to the spin model calculations.\\

\section*{Author contribution}
Z.F., T.K. and L.Y. synthesized and characterized the crystals. Y.H., C.M.G. and P.J.W.M. designed the experiments, performed FIB microstructuring and conducted the transport measurements. Y.H., C.M.G. and P.J.W.M. developed the theoretical framework and performed simulations of the spin and Landau models. All authors were involved in writing the paper.\\

\section*{Competing interests}
The authors declare no competing interests.\\

\newpage
\section*{Supplementary information}

\setcounter{figure}{0}
\renewcommand{\thefigure}{S\arabic{figure}}

\section{Scanning electron microscope images of all \CTS~devices}

Five microstructured devices were fabricated for transport measurements using the procedure described in the Methods, with a bar cross-section of approximately $\mathrm{3\times3 ~\mu m^2}$. Device S1 was patterned in the $ab$ plane; the current was applied along $a$ and the magnetic field along $c$, enabling measurements of the longitudinal response as well as the Hall voltage $V_{ba}$. Devices S2-S5 were patterned in the $ac$ plane; the current was applied along $a$ and the magnetic field along $c$, and only longitudinal voltages were measured. Device S4 corresponds to device S3 after further thinning of the bar cross-section. In device S5, Pt film is to secure the electrical contact between the sample and the Au leads, by FIB-assisted deposition. Unless stated otherwise, the data in the main text ar from device S1, data from the other devices are provided below.

\begin{figure}[H]
    \centering
    	\includegraphics[width = 0.98\linewidth]{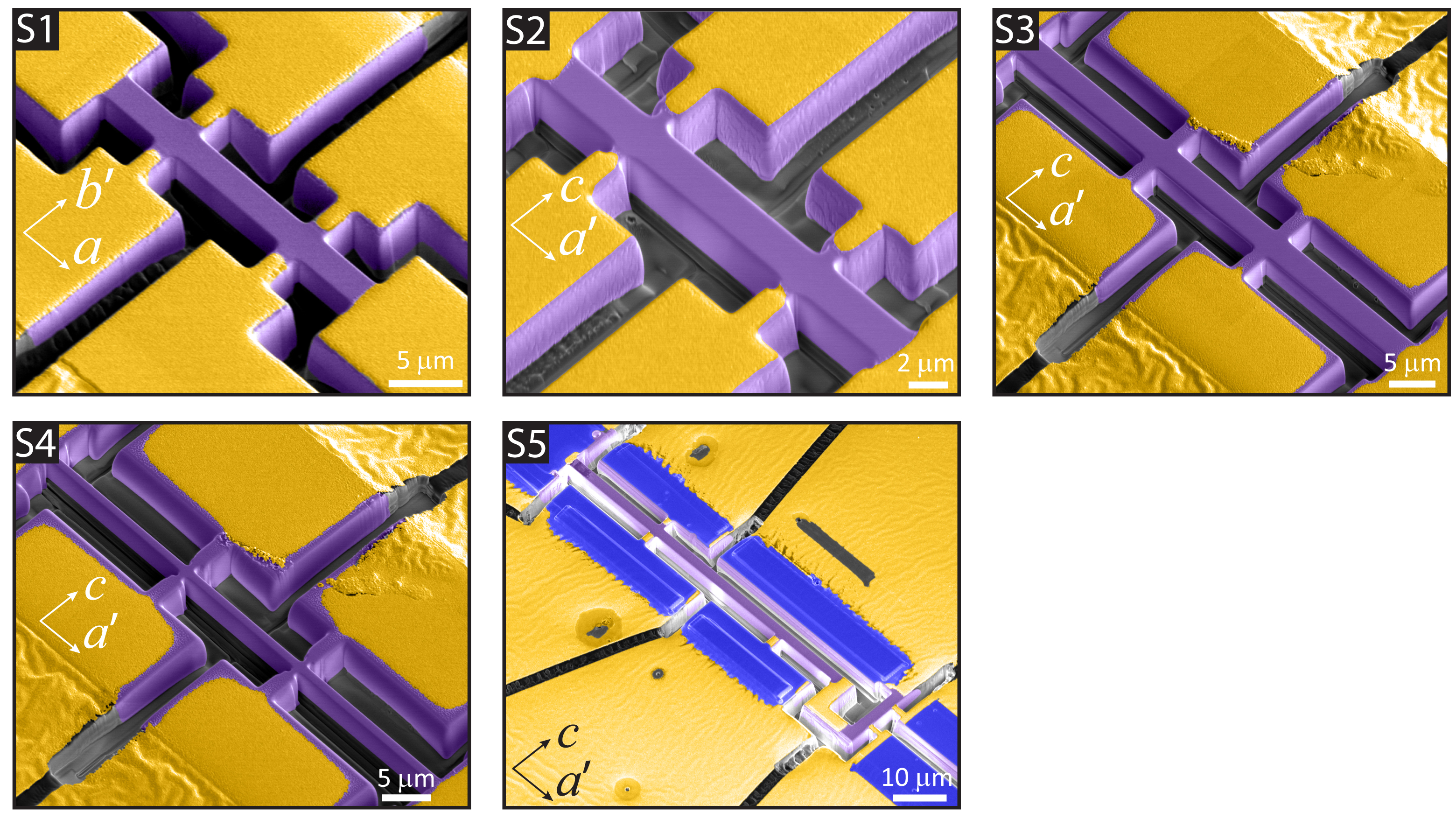}
    \caption{False-color SEM image of the devices. Au-coated contact pads are shown in yellow, the exposed \CTS~region in purple, and FIB-deposited Pt (blue) was used to secure the electrical contact. All microstructures were mounted on glue droplets to mitigate thermal-contraction mismatch and the resulting thermal differential strain at low temperature.}
    \label{figS1}
\end{figure}

\section{Temperature and field dependence of resistivity, eMChA and AHE in all devices}
The resistivity of the five microstructures shows similar temperature dependence, with anomalies at $T_{\mathrm{N1}}\approx 38~\mathrm{K}$ and $T_{\mathrm{N2}}\approx 26.5~\mathrm{K}$, consistent with bulk \CTS~measurements \cite{Park2022}. Microstructure S4 is an exception: the slope changes at $T_{\mathrm{N1}}$ and $T_{\mathrm{N2}}$ are weaker, likely due to its reduced bar thickness, since it was fabricated by thinning S3, decreasing the bar thickness from $\mathrm{3.6~\mu m}$ to $\mathrm{1.4~\mu m}$. This geometric confinement can suppress long-range magnetic order and round the resistive kinks. Aside from this, the overall $\rho(T)$ dependence is consistent across devices, with differences mainly in the residual resistivity. This reproducibility indicates that the observed behaviour is intrinsically inherited from bulk.
\begin{figure}[H]
    \centering
    	\includegraphics[width = 0.5\linewidth]{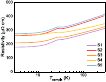}
    \caption{\textbf{Temperature dependence of resistivity in all devices.} Device S1 from crystal \#1, device S2-S4 from crystal \#2, device S5 from crystal \#3. }
    \label{figS2}
\end{figure}
All five devices show similar field dependence under the same measurement geometry (current along the crystalline $a$ axis; magnetic field applied along the $c$ axis).

For $T_{\mathrm{N1}} > T > T_{\mathrm{N2}}$, the magnetoresistance exhibits only the metamagnetic (nematicity) transition: the resistivity decreases smoothly as the $C_3$ rotational symmetry is restored. Below $T_{\mathrm{N2}}$, the transition shifts to lower fields due to coupling to the chiral order, and the magnetoresistance becomes hysteretic. The coercive field $H_c$ corresponds to the chirality-reversal field.

At a fixed temperature (e.g., 12~K), the nematic transition field $H_{\mathrm{nem}}$ is consistent across microstructures, lying slightly below 3~T. By contrast, $H_c$ varies substantially: at 12~K, $H_c \approx 4.5$~T in S1-S3, whereas it is already below 3~T in S5. Device S4 shows a strongly rounded chirality transition, consistent with geometric confinement.

The spread in $H_c$ likely reflects crystal-to-crystal differences, residual strain, and Joule heating. S5 was fabricated from a different single crystal, while S2 and S3 (from the same crystal) exhibit similar $H_c$, suggesting a crystal-dependent contribution. In addition, because the microstructures are mounted on glue rather than fully freestanding, differential thermal contraction between the glue and the microstructure can generate residual strain that shifts $H_c$. A later experiment of microstructured \CTS~mounted on free-standing membranes is worth to try for better understanding the strain effect on $H_c$.  Finally, Joule heating reduces the apparent $H_c$: for S5, increasing the excitation current from 0.3~mA (Fig.~S3) to 1.0~mA (Fig.~S4) lowers $H_c$ by $\sim 1$~T.

In all five devices, the non-nematic state exhibits a lower resistance than the nematic state. This trend is consistent with a preferential alignment of the stripes in the nematic domain along the long axis of the microbar ($Q$ oriented predominantly perpendicular to the bar)\cite{feng2025}. Two effects may contribute to this preference. First, although the Z$_3$ nematic domains are degenerate in an ideal bulk crystal, this degeneracy can be lifted in a sufficiently confined microbar, as a difference in the surface energy between the Z$_3$ nematic domains acts as an effective symmetry-breaking field that biases nematic domain selection. Second, a potential strain arising from differential thermal contraction between the microbar and the underlying glue during cooldown can also couple to the nematic order parameter.

\begin{figure}[H]
    \centering
    	\includegraphics[width = 0.98\linewidth]{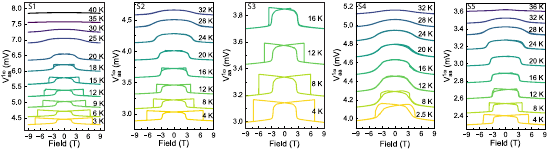}
    \caption{\textbf{Reproducibility of the magnetoresistance in all devices.} Data at different temperatures are vertically offset by 0.2 mV (S1), 0.15 mV (S2, S3) and 0.1 mV (S4, S5) for clarity. Current (density) for each measuremtn: $ \mathrm{1.4~mA~(2\times10^8A/m^2)}$ for S1, $ \mathrm{1.0~mA~(1\times10^8A/m^2)}$ for S2, $ \mathrm{1.0~mA~(9\times10^7A/m^2)}$ for S3, $ \mathrm{0.4~mA~(9\times10^7A/m^2)}$ for S4, $ \mathrm{0.3~mA~(3\times10^7A/m^2)}$ for S5.}
    \label{figS3}
\end{figure}

The second-harmonic signals measured in all five devices reproduce the key signature of eMChA: a hysteretic loop emerges below $T_{\mathrm{N2}}$ as chiral order develops and widens upon cooling. The loop in S4 is again strongly rounded, consistent with geometric confinement. Device S5 exhibits an additional pair of non-antisymmetric kinks within the hysteretic loop, reducing the loop amplitude on the left side and enhancing it on the right. Because this feature lacks antisymmetry, we attribute it to an extrinsic origin, most likely the asymmetric device geometry of S5: the current enters along the $c$ axis, flows along the $a$ axis, and exits along $c$ again. This current path differs from that in S1--S4, for which the current path and associated heating profile are symmetric with respect to the measured bar.

\begin{figure}[H]
    \centering
    	\includegraphics[width = 0.98\linewidth]{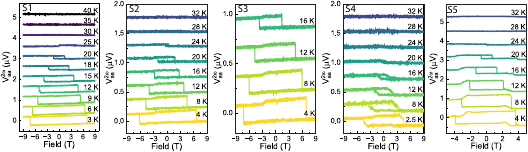}
    \caption{\textbf{Reproducibility of the eMChA in all devices.} Data at different temperatures are vertically offset by 0.5 $\mu$V (S1), 0.25 $\mu$V (S2, S4),  0.3 $\mu$V (S3) and 0.9 $\mu$V (S5) for clarity. Current (density) for each measuremtn: $ \mathrm{1.4~mA~(2\times10^8A/m^2)}$ for S1, $ \mathrm{1.0~mA~(1\times10^8A/m^2)}$ for S2, $ \mathrm{1.0~mA~(9\times10^7A/m^2)}$ for S3, $ \mathrm{0.4~mA~(9\times10^7A/m^2)}$ for S4, $ \mathrm{1.0~mA~(9\times10^7A/m^2)}$ for S5.}
    \label{figS4}
\end{figure}

$V_{ba}$ could be measured only in device S1. It also onsets at $T_{\mathrm{N2}}$ and increases gradually upon cooling.

\begin{figure}[H]
    \centering
    	\includegraphics[width = 0.2\linewidth]{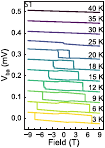}
    \caption{\textbf{AHE of device S1 at different temperatures.} Data at different temperatures are vertically offset by 50 $\mu$V for clarity. }
    \label{figS5}
\end{figure}

\section{Thermal effect on $V_\mathrm{2\omega}$}
Joule heating is a common extrinsic source of higher-harmonic voltage generation. It directly results in an oscillating temperature at a frequency of 2$\omega$ when applying an a.c. electric current I$_{\omega}$. In cases where the contact resistances of the device electrodes are not perfectly balanced, an extrinsic V$_{2\omega}$ can be generated. This is exactly the source of the background subtracted in Fig. 3b. Unlike the intrinsic eMChA signal, the sign of which is flipped by the direction of the magnetic field applied, the extrinsic signal stays nearly field-independent as the thermal imbalance depends mainly on the device geometry rather than the switch of scalar spin chirality. This constant offset was defined as the mean of the two opposite-polarity values measured at zero magnetic field. The corresponding uncorrected data are shown in Fig. S4. 

To quantify the temperature evolution of the offset, we performed zero-field cooling and warming up after training in a $\pm$ 9T magnetic field. As shown in Fig. S6a, the offset remains positive over the measured temperature range. Consistent with a thermal origin, the third-harmonic signal follows a trend proportional to $R\cdot dR/dT$, indicating that Joule heating is non-negligible. The close agreement among the three third-harmonic traces indicates that Joule-heating-related artefacts do not produce hysteresis under our measurement conditions. Therefore, the hysteresis loop observed in the second-harmonic channel cannot be attributed to heating, and therefore reflects a nontrivial eMChA response. 

\begin{figure}[H]
    \centering
    	\includegraphics[width = 0.98\linewidth]{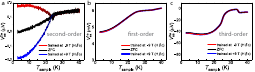}
   \caption{Temperature dependence of the \textbf{a} second harmonics of longitudinal response, \textbf{b} first harmonics of longitudinal response and \textbf{c} third harmonics longitudinal response (from device S1).}
    \label{figS6}
\end{figure}

\begin{figure}[H]
    \centering
    	\includegraphics[width = 0.98\linewidth]{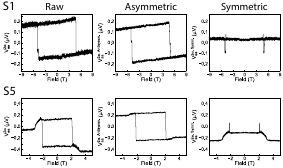}
   \caption{Decomposition of field-asymmetric and symmetric components of the second-harmonics at 12 K in device S1 and S5.}
    \label{figS7}
\end{figure}
This is further elaborated in the analysis of the field-asymmetric/symmetric component of $V_{2\omega}$ signal across different devices. To keep the loop feature in the asymmetric composition, we need to separate the data measured during the up-sweeping and down-sweeping of the field into $V_{\mathrm{sweepUP}}(B)$ and $V_{\mathrm{sweepDOWN}}(B)$, and do calculation as below:
\begin{equation}
    V_{\mathrm{asymm}}(B) = (V_{\mathrm{sweepUP}}(B)-V_{\mathrm{sweepDOWN}}(-B))/2;
\end{equation}
\begin{equation}
    V_{\mathrm{symm}}(B) = (V_{\mathrm{sweepUP}}(B)+V_{\mathrm{sweepDOWN}}(-B))/2.
\end{equation}

While the asymmetric component is nearly identical across devices, the symmetric component differs distinctively between devices S1 and S5. In device S1, it remains vanishingly small across the entire field range; the two sharp peaks at the critical field are most likely attributable to the symmetrizing analysis process rather than to intrinsic behavior. In device S5, however, the symmetric component is clearly observed and exhibits a distinct field dependence. This clear difference between these devices originates directly from their different geometry. The hall-bar-like structure in device S1 ensured a well-defined thermal boundary condition, thereby strongly reducing the thermal gradient across the microstructure. Comparatively, device S5 is shaped into an "L"-like geometry, in which thermal gradient is readily expected due to the asymmetry of the device itself and therefore generates a non-negligible field-symmetric $V_{2\omega}$ component.

\section{Spin model calculations}
To understand how the non-coplanar spin texture evolves under a magnetic field and how this evolution affects the scalar spin chirality and the net magnetization, we construct a spin Hamiltonian and determine the lowest-energy spin configurations using a variational Ansatz for different magnetic fields.

The Hamiltonian is
\begin{equation}
\begin{aligned}
\mathcal{H} =
&\;\frac{J_1}{2}\sum_{\langle i,j\rangle}
\mathbf{S}(\mathbf{r}_i)\cdot\mathbf{S}(\mathbf{r}_j)
+\frac{J_2}{2}\sum_{\langle\!\langle i,j\rangle\!\rangle}
\mathbf{S}(\mathbf{r}_i)\cdot\mathbf{S}(\mathbf{r}_j)
+ \sum_{\langle i,j\rangle}
\mathbf{D}_{ij}\cdot
\left[\mathbf{S}(\mathbf{r}_i)\times\mathbf{S}(\mathbf{r}_j)\right] \\
&+ \frac{B}{2}\sum_{\langle i,j\rangle}
\left[\mathbf{S}(\mathbf{r}_i)\cdot\mathbf{S}(\mathbf{r}_j)\right]^2
+ K\sum_i \left(S_i^{z}\right)^2
- \sum_i H_\mathbf{ext}\cdot\mathbf{S}(\mathbf{r}_i).
\end{aligned}
\end{equation}

where $J_1$ denotes the nearest-neighbor in-plane exchange interaction and sets the overall energy scale, $J_2$ is the next-nearest-neighbor in-plane exchange interaction, $D$ characterizes the strength of the Dzyaloshinskii--Moriya interaction, $B$ is the biquadratic exchange strength, $K$ represents the uniaxial single-ion anisotropy, and $\mathbf{H}_{\mathrm{ext}}$ is the applied external magnetic field.

Because of the translational symmetry of the spin texture, the total energy can be obtained by evaluating the four inequivalent spins within a single magnetic unit cell.\cite{PhysRevMaterials.6.024405} In the energy minimization, the spin amplitudes are fixed, while the polar angles are treated as free variables. The azimuthal angles of the four spins are fixed to $0$, $\pi$, $-\pi/3$, and $\pi/3$, respectively, revealing the fundamental features of the scalar spin chirality.

At zero external field, the energy landscape exhibits two degenerate local minima. One corresponds to the spin configuration shown in the figure, while the other corresponds to its time-reversed counterpart. Upon applying a finite external field, the two minima become non-degenerate, resulting in an energetic preference for one of the two polarities.

To characterize the net magnetization and scalar spin chirality, we evaluate their local distributions on each triangular plaquette formed by three nearest-neighbor spins. The local scalar spin chirality is defined as
\begin{equation}
\chi_{\mathrm{local}} = \mathbf{S}_i \cdot \left( \mathbf{S}_j \times \mathbf{S}_k \right)
\end{equation}
The local net magnetization of each plaquette is defined as
\begin{equation}
S^{z}_{\mathrm{local}} = \frac{S_i^{z} + S_j^{z} + S_k^{z}}{6}
\end{equation}
We find that eight such plaquettes form the basic unit of the local chirality and magnetization distribution pattern. Since transport measurements probe the global state of the device, we calculate the average chirality over the eight plaquettes as
\begin{equation}
\bar{\chi} = \frac{1}{8} \sum \chi_{\mathrm{local}}
\end{equation}
The net magnetization per Co$^{2+}$ ion within this eight-plaquette unit is obtained as
\begin{equation}
 S^{z}/\mathrm{Co^{2+}} = \frac{1}{4} \sum S^{z}_{\mathrm{local}}
\end{equation}

\section{Phenomenological Landau model}

The calculation highlights the role of the coupling $\kappa$ between two symmetry-incompatible magnetic orders.

For $\kappa \neq 0$, the $\phi(B_z)$ response reproduces the key experimental trends in the magnetoresistance: a negative magnetoresistance below $T_{\mathrm{N1}}$ and hysteresis emerging below $T_{\mathrm{N2}}$. In this regime, switching the spin chirality simultaneously suppresses the nematic order, leading to an abrupt reduction in the magnetoresistive signal. The zero-field resistivity is identical for opposite chiralities.

For $\kappa = 0$, $\chi(B_z)$ is unchanged, whereas $\phi(B_z)$ differs qualitatively below $T_{\mathrm{N2}}$. Because the nematic AFM and chiral AFM orders are decoupled, the nematic switching field $H_{\mathrm{meta}}$ is not reduced by the presence of chiral order $\chi$ and can continue to increase on cooling as $H_{meta} \propto  (T_{N1}-T)$. Consequently, the chirality can no longer be inferred from the resistance.
\begin{figure}[H]
    \centering
    	\includegraphics[width = 0.8\linewidth]{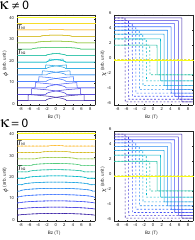}
    \caption{Calculated nematic order parameter $\phi$ and chiral order parameter $\chi$ with and without the field induced coupling term $\kappa$, as functions of magnetic field for temperatures between 2 K and 41 K, with step of 3 K. Parameters used: $H_{c,0} = 7, \alpha = 1, b_\phi = 30, a_\chi/b_\chi = 1.3, H_m = 1.15, a_\phi = 2.3, \beta_\phi = 0.95$ and $\kappa = 5.5$ or $0$.}
    \label{figS8}
\end{figure}
The details of the Landau model are as belows.

A phenomenological Landau model captures some of the essential physics of coupling a nematic and a chiral order parameter. A salient assumption is the coexistence of both order parameters within the spin-texture of \CTS, which may be achieved via asymmetric multi-Q orders. Like any Landau model, it serves as a controlled sandbox illustrating how two symmetry-incompatible magnetic orders, when coupled, can exchange their experimental signatures and generate qualitatively new switching phenomena. As such, the model should be understood as a lowest-order effective description that isolates the essential ingredients required to reproduce the experimentally observed sequence of phase stabilization, collapse, and hysteretic readout. 

A full model must treat both order parameters, the nematic order  $\phi$ and the chiral order $\chi$, on equal footing and self-consistently solve a coupled model for both. However, the pronounced hysteresis and important strong first-order nature suggest that dynamical effects are key for any chirality switching behavior, such as domain motion and unpinning. This problem may be addressed by time-dependent Ginzburg-Landau model, provided in-depth knowledge about the nature of the coupled domains and their energetics. To gain insights into the essential physics, instead we focus on a static model. The order parameter $\chi$ will be introduced as a phenomenological field based on the data, in particular endowed with manual hysteresis based on the experimental observations. Its two enantiomeric states $\chi=\pm\chi_0$ correspond to opposite scalar spin chiralities. Importantly, $\chi$ is treated as a discrete order parameter with a fixed magnitude $\chi_0(T)$ below the chiral ordering temperature $T_{N2}$, reflecting the experimentally observed abrupt switching behavior:
\begin{equation}
\chi_0 = \sqrt{(T_{N2}-T)a_\chi/b_\chi}
\end{equation}
This allows to formulate a static Landau model for the nematic order $\phi$, and to evaluate its modifications based on the coupling and sign-change of the chiral order.

Following this spirit, one may write the following Free energy expansion:

\begin{equation}
\mathcal{F}
=
\frac{a_\chi}{2}\,(T-T_{N2})\,\chi^2
+\frac{b_\chi}{4}\,\chi^4
+\frac{a_\phi}{2}\,(T-T_{N1})\,\phi^2
+\frac{b_\phi}{4}\,\phi^4 
+\frac{\beta_\phi}{2}\left(\frac{H_z}{H_m}\right)^2 \phi^2
-\frac{\kappa}{2}\,H_z\,\chi\,\phi^2
.
\end{equation}

This model describes two distinct magnetic orders that coexist at low temperatures, where $a_i$ and $b_i$ denote the Landau parameters yielding separate transitions of the chiral order $\chi$ at $T_{N2}$ as well as the nematic $\phi$ at $T_{N1}$ in the absence of coupling. The nematic order parameter $\phi$ is suppressed by out-of-plane magnetic fields $H_z$, which is captured via a metamagnetic field scale $H_m$ and the associated Landau parameter $\beta_\phi$. In the absence of coupling, this term suppresses nematicity at a metamagnetic field self-consistently as 
\begin{equation}
H_{\mathrm{meta}}(T)=H_m\sqrt{\frac{a_\phi\,(T_{N1}-T)}{\beta_\phi}}. 
\end{equation}

By construction, the two order parameters belong to different symmetry channels. In zero magnetic field they are symmetry-incompatible and may emerge independently at distinct transition temperatures. A magnetic field applied along the $c$ axis introduces a time-reversal-odd axial vector, which allows additional couplings between $\chi$ and $N$ that are otherwise forbidden. These are encoded in the last term, $\frac{\kappa}{2}\,H_z\,\chi\,\phi^2$, which encodes a field-induced coupling between the chiral and nematic orders, tuned by a Landau parameter $\kappa$. Critically, the product $H_z \chi$ is even under time-reversal symmetry, yet still sensitive to the sign of $\chi$.

Given that only $\phi$ will be solved for, the effect of finite $\chi$ can be considered a renormalization of the Landau parameters. This allows to simplify the Landau free energy for the nematic order,
\begin{equation}
F_\phi(\phi)=\frac{1}{2}a_\phi^{\mathrm{eff}}(T,H,\chi)\,\phi^2+\frac{b_\phi}{4}\phi^4,
\end{equation}
with $b_\phi>0$ ensuring stability. The effective quadratic coefficient is simply given by
\begin{equation}
a_\phi^{\mathrm{eff}}(T,H,\chi)=a_\phi(T-T_{N1})+\beta_\phi(H_z/H_m)^2-\kappa\,H_z\chi
\end{equation}

This form of coupling is central to the physics captured by the model. Unlike a linear coupling proportional to $\phi$, which would merely select the orientation of nematic domains, the quadratic coupling modifies the stability of the nematic phase itself. Depending on the sign of $\kappa$ and the chirality $\chi$, the effective coefficient $a_\phi^{\mathrm{eff}}$ can be either reduced or enhanced. As a result, one chiral enantiomer stabilizes nematic order to higher fields than expected from $a_\phi$ alone, while the opposite enantiomer destabilizes it. This mechanism allows the nematic phase to persist metastably beyond its nominal critical field and to collapse abruptly when $\chi$ reverses.

Minimization of $F_\phi$ yields a simple amplitude response,
\begin{equation}
\phi=
\begin{cases}
0, & a_\phi^{\mathrm{eff}}\ge 0,\\[4pt]
\sqrt{-a_\phi^{\mathrm{eff}}/b_\phi}, & a_\phi^{\mathrm{eff}}<0,
\end{cases}
\end{equation}
which is continuous in $a_\phi^{\mathrm{eff}}$ but can nevertheless display a sharp drop when $a_\phi^{\mathrm{eff}}$ jumps discontinuously due to a reversal of $\chi$ at fixed field.

The hysteretic behavior of the chiral order $\chi$ is incorporated phenomenologically through a pinned-Ising switching rule. The sign of $\chi$ is allowed to change only when the applied magnetic field exceeds a temperature-dependent coercive field $H_c(T)$. This coercive field encodes domain pinning and slow dynamics that are not treated explicitly in the static free-energy functional. $H_c(T)$ is chosen to match the experimentally observed hysteresis, and accordingly to vanish continuously at the chiral ordering temperature $T_{N2}$,
\begin{equation}
H_c(T)=H_{c,0}\left(\dfrac{T_{N2}-T}{T_{N2}}\right)^{\alpha}, \quad T<T_{N2}
\end{equation}

where $H_{c,0}$ is the zero-temperature limit coercive field and $\alpha$ controls the functional form. This choice ensures that the hysteresis loop in $\chi(H)$ closes continuously at $T_{N2}$. The precise functional form is not essential but merely provides a smooth interpolation between experimentally established limits.

Within this framework, the coupled switching behavior follows directly by construction. At a fixed temperature below $T_{N2}$, the magnetic field $H_z$ is swept over the entire field range, completing a hysteresis cycle. For one chirality, the quadratic coupling reduces $a_\phi^{\mathrm{eff}}$, allowing nematic order to survive well beyond the field scale $H_{\mathrm{meta}}(T)$ at which it would otherwise disappear. When the coercive field is reached and $\chi$ reverses, the sign of the coupling term changes abruptly. If the parameters are such that $a_\phi^{\mathrm{eff}}$ becomes positive after the flip, the nematic order collapses immediately to $N=0$. Upon further increase of the field, nematicity remains absent. On the reverse field sweep, the nematic phase reappears only once $a_\phi^{\mathrm{eff}}$ becomes negative again despite the destabilizing chirality of $\chi$ at a given field polarity, typically near the metamagnetic scale. Importantly, as the coupling coefficient is proportional to $H_z$, once the field is swept through zero into the negative quadrant, its sign change reverses the role of stabilizing/destabilizing chirality, thus completing a copy of the hysteretic cycle at negative polarity.

The present poor-mans model intentionally omits many aspects that would be required for a quantitative or microscopic theory. In particular, it does not describe the internal structure of magnetic domains, the dynamics of domain-wall motion, or the microscopic origin of the coupling between chirality and nematicity. It also treats the nematic transition as continuous in the Landau sense, while in fact $\phi$ may itself display genuine first-order behavior and metastability at low temperature.  the inclusion of higher-order terms or explicit pinning rules for the nematic order itself.

Nevertheless, the model captures, at the lowest effective level, how symmetry-allowed coupling between two magnetic orders can qualitatively reshape their phase diagram and experimental signatures. In this sense, it provides a transparent framework for understanding how information stored in a topological antiferromagnetic state can be transferred to, and read out through, a coexisting nematic order.

\bigskip






\end{document}